\documentclass[aps,graphicx,twocolumn]{revtex4}%
\usepackage{amsmath}
\usepackage{graphicx}
\begin{document}
%\begin{CJK*}{GBK}{}

\title{Error-rejecting quantum computing with solid-state spins assisted by
low-Q optical microcavities\footnote{Published in Phys. Rev. A
\textbf{94}, 062310 (2016)}}

\author{Tao Li and  Fu-Guo Deng\footnote{Corresponding author: fgdeng@bnu.edu.cn}}

\affiliation{Department of Physics, Applied Optics Beijing Area
Major Laboratory, Beijing normal University, Beijing 100875, China}

\date{\today }

\begin{abstract}
We present an efficient proposal for error-rejecting quantum
computing with quantum dots (QD) embedded in single-sided optical
microcavities based on the interface between the circularly
polarized photon and  QDs. An almost unity fidelity of the quantum
entangling gate (EG) can be implemented with a detectable error that
leads to a recycling EG procedure, which improves further the
efficiency of our proposal  along with the robustness to the errors
involved in imperfect input-output processes. Meanwhile, we discuss
the performance of  our proposal for the EG on two solid-state spins
with currently achieved experiment parameters, showing that it is
feasible with current experimental technology. It provides a
promising building block for solid-state quantum computing and
quantum networks.

\end{abstract}
%\pacs{03.67.Lx, 03.67.Pp, 42.50.Ex, 42.50.Pq, 03.67.Hk}
\maketitle
%\end{CJK*}

\section{Introduction}

Compared with the traditional computer, quantum computing \cite{QIP}
can factor an $n$-bit integer with the magical Shor  algorithm
\cite{Shor}, exponentially faster than the best-known classical
algorithms. It can also run the famous quantum search algorithm, the
Grover algorithm \cite{Grover} or the optimal Long  algorithm
\cite{Long},  for unsorted database search, which requires
O($\sqrt{N}$) operations only, rather than O($N$) operations
involved in its classical counterpart. Both  circuit-based quantum
computing and the measurement-based one require quantum entangling
gates. That is,  the ability to entangle the quantum bits (qubits)
is an essential building block in the construction of a quantum
computer \cite{QIP}. Since the early quantum entangling gate (EG)
for single atoms was designed with the assistance of a high-Q
optical cavity \cite{EGatoms1994}, more and more attention has been
paid to the entangling operation between stationary qubits
\cite{EGatoms1994,EGatoms1997,EGDI2000,EGatomsZheng,EGatomsExp,EGatomsHerelded,Cex1,Cex2,Cex3}.

The previous  EG between two stationary qubits is implemented by
various methods that resort to different interactions, i.e., the
coherent control of the  direct  qubit-qubit interaction, the
indirect interaction meditated with high-Q optical cavities
\cite{EGatoms1994,EGatoms1997,EGDI2000,EGatomsZheng,EGatomsExp,EGatomsHerelded},
or the controllable exchange interaction involved in the solid-state
spin systems \cite{Cex1,Cex2,Cex3}. The typical absence of a
heralding measurement in the EG resulting from the  direct or
indirect qubit-qubit interaction will lead to some ambiguous error,
such as the one originating from the photon loss as a result of the
cavity decay or the radiative deexcitation of the stationary dipole.
These proposals could work successfully under the condition that the
amount of noise involved in these EGs is less than a small threshold
value \cite{Threshhold1}. It will be more physical-resource
consuming and largely increase the complexity of the target quantum
system when performing  scalable quantum computing with  EGs of a
higher error probability \cite{Threshhold2,Threshhold3}.  However,
with a layered quantum-computer architecture,  the resources
required for error correction will become manageable when the
physical error rate is about an order of magnitude below the
threshold value  of the chosen code \cite{Threshhold3}.

An alternative strategy exploits a measurement on the auxiliary
photonic qubits that entangle  with the corresponding stationary
qubits to project the target stationary system into an entangled
state, which constitutes a quantum EG of  high fidelity. Meanwhile,
its success is heralded by the detection of  photons
\cite{EGInter1,EGInter2,EGInter3,EGInter4,EGInter5,EGInter6,EGInter7,EGInter8,EGInterExp}.
Its fidelity does not suffer from the photon loss noise, and it is
relatively robust to the variation of the system parameters. Since
these special schemes involve the optical Bell-state measurement
(BSM) assisted by linear optical elements, they can succeed with the
maximal efficiency of $1/2$ in the ideal situation
\cite{BSA,ClusterLinear}. Besides the nondeterministic efficiency,
the two photons for the BSM are required to be indistinguishable in
all degrees of freedom except for the one used to encode the quantum
information \cite{heraldedm}. Therefore, if the polarization degree
of freedom of photons is used to encode the photonic qubit, it will
result in a smooth degradation in the performance of these EGs when
the photonic spectral differences and the practical experimental
control of the arrival time of the photons  are considered.

It is well known that solid-state spin  systems offer a promising
candidate  for the realization of quantum computing
\cite{Hybridatom,ClusterQD3,QCspins}. Its solid-state nature
combined with nanofabrication techniques provide  a relatively
simple way to incorporate the spins into optical microcavities and
allow  for the generation of arrays of solid qubits
\cite{NVQED,QDQED,QDspins,hybridspins}. One attractive type of
solid-state spin system is the electron spin in a quantum dot (QD)
\cite{QDspins,QDspins1,QDspins2}. Not only does it provide easy ways
of optical initialization, single-qubit manipulation, and  readout,
but also it processes a long coherence time of the electron spin in
QDs, which is typically around several microseconds  when spin echo
techniques are used \cite{qdop,time,qd26,QDCspins}. Most existing
quantum computing schemes based on single photons and single spins
in QDs are performed in a strong coupling regime as a result of
cavity quantum electrodynamics (QED)
\cite{QDspins,EGQdHu1,EGQdHu2,QCwei1,QCwei2}. However, the strong
coupling regime remains a challenge and some  EG proposals for QDs
in a low-Q cavity are proposed at the price of a decrease in the
fidelity and the efficiency of the EG
\cite{QDweak1,QDweak2,QDweak3,EGQdKoshino,EGQdYoung}. Meanwhile, the
solid-state spins used in  quantum computing are supposed to be
homogeneous and the inhomogeneity of the spins will further decrease
the feasibility of the proposed solid-state spin quantum computing
\cite{ClusterQD3}.

In this article, we propose a robust proposal for the quantum EG
between two QDs embedded in single-sided microcavities
\cite{EGQdHu1}. It is a practical proposal for performing efficient
solid-state quantum computing that overcomes the existing
limitations, since the fidelity of the EG for two QDs is always
towards unity and the efficiency of the EG can, in principle, also
approach unity. Compared with previous EGs for QDs based on cavity
QED \cite{EGQdHu2,QCwei1,QCwei2,QDweak1,EGQdYoung,EGQdKoshino}, the
present scheme also has several other advantages. First, it does not
require the strong-coupling limit and can work efficiently in low-Q
cavities or even in the regime of resonance scattering
\cite{EGQdYoung} where the modified spontaneous emission parameter
of QDs coupled  resonantly to a microcavity is matched to that in
the bulk dielectric.  Second, the success of our  EG  is heralded
with fidelity larger than $0.99$, and it is signaled by the
detection of a photon of orthogonal polarization as a result of
cavity QED \cite{adjustBS0}, where only one effective input-output
process is involved in single-sided cavities, similar to the one in
a nitrogen-vacancy center coupled to two-sided cavities
\cite{NVNemoto}. This is the origin of our high efficiency rather
than the maximal efficiency $1/8$ based on two-sided cavities.
Third, the imperfect reflection of the cavity due to the deviation
from the ideal conditions, i.e., the nonzero photonic bandwidth, the
finite coupling rate between the QD and cavity mode, and the
mismatch between the incident photon and the cavity mode, can only
lead to photon loss or  a click on either vertical detector  rather
than a decrease in the fidelity of the EG. Meanwhile, the QD
subsystem will be collapsed into the original state when one of
vertical detectors clicks, and we can input another probe photon to
restart the EG directly without any re-preparation of the QDs, which
makes our proposal more efficient than others. With our EGs, one can
implement universal quantum computing, including both the one-way
quantum computing and the circuit one  \cite{QIP}.

\section{Error-rejecting entangling gate for two QDs in low-Q optical microcavities}

\begin{figure}[!tpb]%[tpb]           %Figure 1
\begin{center}
\includegraphics[width=8.025 cm,angle=0]{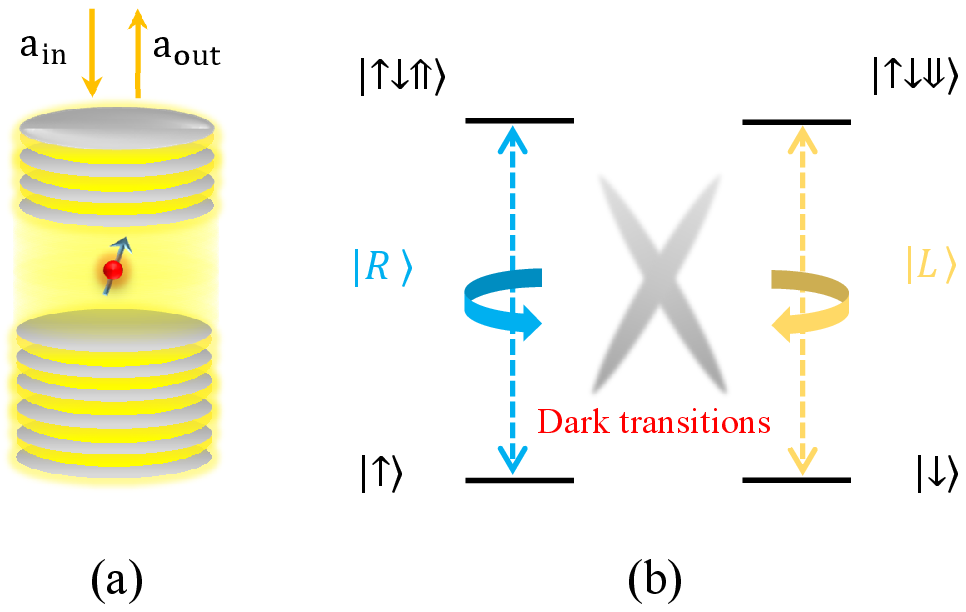}
\caption{The spin-dependent transitions for negatively charged
exciton $X^-$. (a) A singly charged QD inside a single-sided optical
micropillar cavity. (b) The relative energy levels and the optical
transitions of a QD.} \label{fig1}
\end{center}
\end{figure}

Let us consider a quantum system consisting of a singly charged
self-assembled In(Ga)As QD embedded in a single-sided micropillar
cavity \cite{QDQED,EGQdHu1,EGQdHu2,QCwei1}.   The quantization axis
$z$ is chosen along the growth direction of the QD and is also
parallel to the light propagation direction, shown in
Fig.~\ref{fig1}(a). The dipole transition associated with the
negatively charged QD is strictly governed by Pauli's exclusion
principle \cite{EGQdRev}, shown in Fig.~\ref{fig1}(b).  The single
electron ground states have $J_z=\pm1/2$, denoted
$|\!\uparrow\rangle$ and $|\!\downarrow\rangle$, respectively, and
the optical excited states are the trion states
($X^-=\{|\!\!\uparrow\downarrow\Uparrow\rangle$ or
$|\!\!\uparrow\downarrow\Downarrow\rangle\}$) consisting of two
antisymmetric electrons in the singlet state
$1/\sqrt{2}(|\!\uparrow\downarrow\rangle-|\!\downarrow\uparrow\rangle)$
and one hole with $J_z=\pm3/2$ ($|\!\Uparrow\rangle$ and
$|\!\Downarrow\rangle$). The dipole-allowed transitions  between the
ground state and the trion state are
$|\!\uparrow\rangle\leftrightarrow|\!\uparrow\downarrow\Uparrow\rangle$
and
$|\!\downarrow\rangle\leftrightarrow|\!\uparrow\downarrow\Downarrow\rangle$,
along with the absorbtion of a right-handed circularly polarized
photon $|R\rangle$ and a left-handed one $|L\rangle$, respectively,
while the crossing transitions are dipole forbidden \cite{EGQdRev}.

When a circularly polarized probe photon is launched into the
single-sided cavity, it will be reflected by the cavity with a
spin-dependent reflection coefficient $r_j(\omega)$
\cite{QDQED,EGQdHu1,EGQdHu2,QCwei1}. The dynamic process can be
represented by Heisenberg equations for the cavity field operator
$\hat{a}$ and dipole operator $\hat{\sigma}_-$ in the interaction
picture \cite{book2},
\begin{equation}                           \label{eq1}     % Eq_1
\begin{split}
\frac{d \hat{a}}{dt}   &=   -\left[i(\omega_c\!-\!\omega) \!+\!
\frac{\kappa}{2}\!+\!\frac{\kappa_s}{2}\right] \hat{a}
-g\hat{\sigma}_{-} \!-\!\sqrt{\kappa}\, \hat{a}_{in}\!+\!\hat{R},  \\
\frac{d\hat{\sigma}_-}{dt}   &=   -\left[i(\omega_{X^-}\!-\! \omega)
\!+\!
\frac{\gamma}{2}\right] \hat{\sigma}_{-} \!-\! g\hat{\sigma}_z \, \hat{a}  +\!\hat{N},%\\
 \end{split}
\end{equation}
where $\omega_{X^-}$, $\omega_{c}$, and $\omega$ are the frequencies
of  the dipole transition, the cavity resonance, and the probe
photon, respectively. $\hat{R}$  and $\hat{N}$ are noise operators
which help to preserve the desired commutation relations. The
parameter  $g$ is the coupling strength between $X^-$ and the cavity
mode. $\kappa$ describes the coupling to the input and output ports,
while $\kappa_s$  and $\gamma$ represent the cavity leakage rate and
the trion $X^-$ decay rate, respectively. In the weak excitation
limit where the QD dominantly occupies the ground state,  assisted
by the standard cavity input-output theory $\hat{a}_{out}  =
\hat{a}_{in} \!+\!\sqrt{\kappa}\, \hat{a}$ \cite{book2}, one can
obtain the spin-dependent reflection coefficient
\cite{EGQdHu1,EGQdHu2,EGQdReflection,outputAn}:
\begin{eqnarray}    % Eq_2
r_j(\omega)=1\!-\!\frac{\kappa\left[i(\omega_{X^-}-\omega)+\frac{\gamma}{2}\right]}
{\left[i(\omega_{X^-}-\omega)
+\frac{\gamma}{2}\right]\!\!\!\left[i(\omega_{c}\!-\!\omega)\!+\!\frac{\kappa}{2}\!+\!\frac{\kappa_s}{2}\right]\!+\!jg^2}.\nonumber\\
\label{rcoe}
\end{eqnarray}
Here the subscript $j$ is used to discriminate the case that the
polarized probe photon agrees with the trion transition ($j=1$) and
feels a QD-cavity coupled system  and the case that the polarized photon
decouples from the trion transition ($j=0$) and feels an empty
cavity.

Suppose the electron spin $s$ of a QD is initialized to
$|\psi_s\rangle=\alpha|\!\uparrow\rangle_s
+\beta|\!\downarrow\rangle_s$, with $|\alpha|^2+|\beta|^2=1$. When
the input photon is in the polarized state
$|\psi_p\rangle=\frac{1}{\sqrt{2}}(|R\rangle_{p}-|L\rangle_{p})$,
the photon reflected by the cavity directly due to the mismatching
between the incident probe photonic field and the cavity mode, or
reflected by the desired cavity-QD system, together with the QD,
evolves into an unnormalized state
\begin{eqnarray}   % Eq_3
|\Phi\rangle_{H}\!\!&=&\!\!\frac{\eta_{in}}{\sqrt{2}}\Big[(r_1\times\alpha|\!\uparrow\rangle_s
+r_0\times\beta|\!\downarrow\rangle_s)\otimes|R\rangle_{p}\nonumber\\
&&\!\!-(r_0\times\alpha|\!\uparrow\rangle_s
+r_1\times\beta|\!\downarrow\rangle_s)\otimes|L\rangle_{p}\Big]\;\;\;\;\;\;\;\;\;\;\;\;\;\;\nonumber\\
&&\!\!+\sqrt{1-\eta^2_{in}}\,|\psi_s\rangle\otimes|\psi_p\rangle.
\end{eqnarray}
Here $\eta_{in}$ is the probability amplitude of the photon
reflected by the desired  cavity-QD system
\cite{cavitypillarcoupling}. If one rewrites $|\Phi\rangle_{H}$ with
the linear-polarization basis $\{|H\rangle \equiv
\frac{1}{\sqrt{2}}(|R\rangle+|L\rangle)$, $|V\rangle \equiv
\frac{1}{\sqrt{2}}(|R\rangle-|L\rangle)\}$, one can get the system
composed of the photon \emph{p} and the electron spin $s$ evolving
into a partially entangled hybrid state,
\begin{eqnarray}   % Eq_4
|\Phi\rangle_{H_0}\!\!&=&\!\!\! \left[\frac{\eta_{in}}{2}
(r_1\!+\!r_0)
\!+\!\sqrt{1\!-\!\eta^2_{in}}\,\right]\!\!(\alpha|\!\uparrow\rangle+\beta|\!\downarrow\rangle)_s\otimes|V\rangle_{p} \nonumber\\
&&\!\!\!+\frac{\eta_{in}(r_1-r_0)}{2}
(\alpha|\!\uparrow\rangle-\beta|\!\downarrow\rangle)_s\otimes|H\rangle_{p}.\;\;\;\;\;\;\;\;
\label{eqfaithful}
\end{eqnarray}
Here the photon \emph{p} is partially entangled with the electron
spin $s$, and one can determine the state of the spin according to
the outcome of the measurement on photon \emph{p}. In detail, the
detection of an $|H\rangle_{p}$ photon leads to a phase-flip
operation on spin $s$. Alternatively, the detection of a
$|V\rangle_{p}$ photon signals an error and results in an unchanged
electron spin $s$, no matter where the error originates (the
mismatch  between the incident field and the cavity mode, the low-Q
cavity, or the detuning). For simplicity, we can take
$\eta_{in}\equiv1$ below, since it will not affect the dominant
performance of our EG protocol, and can only reduce the efficiency
of our protocol by the amount of $1-\eta_{in}^2$. Meanwhile, the
output state of the combined hybrid system composed of the spin $s$
and the probe photon $p$ only depends on the combined coefficients
$r_1-r_0$ or $r_1+r_0$ of the cavity-QD system, while it is
independent of the particular parameters that affect the reflection
coefficients $r_j$, shown in Eq. (\ref{rcoe}). Therefore, the output
states of two individual inhomogeneous electron spins embedded in
different optical microcavities  along with their respective probe
photons could be amended to be the same one by utilizing an
adjustable beam splitter \cite{adjustBS0,adjustBS}. The negative
effect of the inhomogeneity of the solid-state spins could be
eliminated formally, which leads to the same result as in
homogeneous cavity-QD systems \cite{QDQED,EGQdHu1,EGQdHu2,QCwei1}.

\begin{figure}[!tpb]%[tpb]           %Figure 2
\begin{center}
\includegraphics[width=8.025 cm,angle=0]{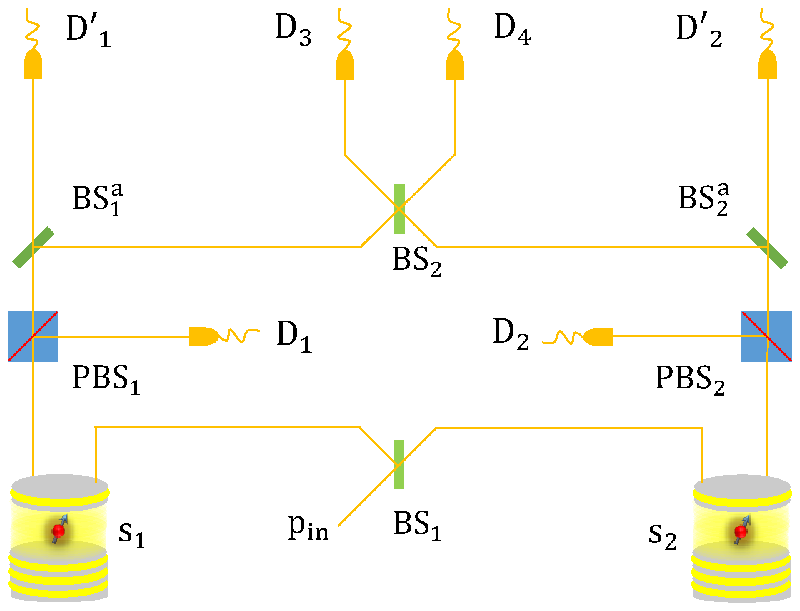}
\caption{The schematic setup of the EG. BS represents a $50:50$ beam
splitter. PBS is the polarizing beam splitter that transmits
$|H\rangle$ photons and reflects $|V\rangle$  photons. BS$^a_i$
denotes the beam splitter with adjustable reflection coefficient
$r^a_i$, i.e., $r^a_1=r^a_2=1$ is utilized for two identical
cavity-QD systems; otherwise, $|r^a_1 r^a_2|<1$ that might lead to
the click of single-photon detector $D'_i$ and restart the recycling
procedure before a phase-flip operation on spin $s_i$.} \label{fig2}
\end{center}
\end{figure}

With the faithful process described above, we can construct an
error-rejecting EG, shown in Fig.~\ref{fig2}, for two identical
electron spins $s_1$ and $s_2$ (the reflection coefficients
$r_i^a=1$ of the adjustable beam splitter BS{}$^a_i$ are adopted),
which will collapse spins $s_1$ and $s_2$ into a state with a
deterministic parity after the entangling process. Suppose the
electron $s_i$ ($i=1,2$) is initially in the state
$|\Phi\rangle_{s_i}=\alpha_i|\!\uparrow\rangle_{s_i}+\beta_i|\!\downarrow\rangle_{s_i}$
with $|\alpha_i|^2+|\beta_i|^2=1$. One probe photon \emph{p} in
state
$|\Phi\rangle_p=\frac{1}{\sqrt{2}}(|R\rangle_{p}-|L\rangle_{p})$
launched into the import of the EG passes through the beam splitter
(BS$_1$),  and it will be reflected by either the left cavity
containing the electron spin $s_1$ or the right one containing
$s_2$. The unnormalized state of the hybrid system composed of the
photon \emph{p} and the electron spins $s_1$ and $s_2$ after being
reflected by the cavities evolves into
\begin{eqnarray}   % Eq_5
|\Phi\rangle_{H_1}\!\!&=&\!\!\frac{1}{\sqrt{2}}\Big\{\!
(r_1\!+\!r_0)(\alpha_1|\!\uparrow\rangle
\!+\!\beta_1|\!\downarrow\rangle)_{s_1} (\alpha_2|\!\uparrow\rangle
\!+\!\beta_2|\!\downarrow\rangle)_{s_2} \;\;\;\;\;\;\nonumber\\
&&\!\! \otimes(|V\rangle_{p_1}\!+\!|V\rangle_{p_2})
\!+\!(r_1\!-\!r_0)\big[(\alpha_1|\!\uparrow\rangle\!-\!\beta_1|\!\downarrow\rangle)_{s_1}\nonumber\\
&&\!\!
\otimes(\alpha_2|\!\uparrow\rangle\!+\!\beta_2|\!\downarrow\rangle)_{s_2}|H\rangle_{p_1}
\!+\!(\alpha_1|\!\uparrow\rangle\!+\!\beta_1|\!\downarrow\rangle)_{s_1}\nonumber\\
&&\!\! \otimes(\alpha_2|\!\uparrow\rangle
\!-\!\beta_2|\!\downarrow\rangle)_{s_2}|H\rangle_{p_2}\big]\Big\}.
\label{Removeerror}
\end{eqnarray}
Here the subscripts $p_1$ and $p_2$ denote photon components that
occupy the left path and the right path, respectively. When the
photon is in the horizonal polarized  state $|H\rangle_{p_1}$ or
$|H\rangle_{p_2}$, the two different spatial modes of photon
\emph{p} are combined on the BS$_2$. The interference of
$|H\rangle_{p_1}$ and $|H\rangle_{p_2}$ modes will collapse  the
hybrid system into
\begin{eqnarray}    % Eq_6
|\Phi\rangle_{H_2}\!\!&=&\!\!\frac{1}{2}(r_1\!-\!r_0)
\big[\!(\alpha_1\alpha_2|\!\uparrow\rangle_{s_1}|\!\uparrow\rangle_{s_2}
\!\!\!-\!\beta_1\beta_2|\!\downarrow\rangle_{s_1}|\!\downarrow\rangle_{s_2}\!)
|\!H\rangle_{p_3}\nonumber\\
&&\!\!+
(\alpha_1\beta_2|\!\uparrow\rangle_{s_1}|\!\downarrow\rangle_{s_2}
\!\!\!-\!\beta_1\alpha_2|\!\downarrow\rangle_{s_1}
|\!\uparrow\rangle_{s_2}\!)|\!H\rangle_{p_4}\big].
\label{pcdfaith}
\end{eqnarray}

Upon a click of the detector $D_3$ or $D_4$, the EG is completed and
the electron-spin system $s_1s_2$ is  projected into a subspace with
a deterministic parity, which is independent of the reflection
coefficients $r_j$, since $r_j$ only appears as a global coefficient
in Eq. (\ref{pcdfaith}).  In detail, when the photon  detector $D_3$
clicks, the spins $s_1s_2$ collapse into the even-parity entangled
state of the form
\begin{eqnarray}    % Eq_7
|\Phi\rangle_{E}\;=\;\alpha_1\alpha_2|\!\uparrow\rangle_{s_1}|\!\uparrow\rangle_{s_2}
-\beta_1\beta_2|\!\downarrow\rangle_{s_1}|\!\downarrow\rangle_{s_2}.
\label{pcdE}
\end{eqnarray}
When the photon detector $D_4$  clicks, the spins $s_1s_2$  are
projected into the odd-parity entangled state of the following form
\begin{eqnarray}     % Eq_8
|\Phi\rangle_{O}\;=\;\alpha_1\beta_2|\!\uparrow\rangle_{s_1}|\!\downarrow\rangle_{s_2}
-\beta_1\alpha_2|\!\downarrow\rangle_{s_1}|\!\uparrow\rangle_{s_2}.
\label{pcdfaith0}
\end{eqnarray}
Both states $|\Phi\rangle_{E}$ and $|\Phi\rangle_{O}$ keep the
information of  the initial state.  Therefore, the coefficient
$\alpha_i$ and  $\beta_j$ could be the  state of   other QD spins
that are entangled with $s_1$ and  $s_2$, which makes the  EG
effective for constructing cluster states in the next section. The
total probability that either $D_3$ or $D_4$  detects one photon of
horizonal polarization  is $\eta_{_H}$:
\begin{eqnarray}    % Eq_9
\eta_{_H}=\frac{|r_1-r_0|^2}{4}. \label{etav}
\end{eqnarray}
Here $\eta_{_H}$ equals  the efficiency of the EG without recycling
procedure.

The first term on the right-hand side of  Eq.  (\ref{Removeerror})
contains the vertical polarization component  $|V\rangle_{p_1}$
($|V\rangle_{p_2}$)  and it will lead to a click on the photon
detector $D_1$ ($D_2$).  In this time, the state of the electron
spins $s_1 s_2$  is projected into
$|\Phi\rangle_{s_1}\otimes|\Phi\rangle_{s_2}$, exactly identical to
the original one without any interaction between the spins and the
photon \emph{ p}, which takes place with  probability $\eta_{_V}$:
\begin{eqnarray}   % Eq_10
\eta_{_V}=\frac{|r_1+r_0|^2}{4}. \label{etah}
\end{eqnarray}
Here $\eta_{_V}$ equals  the heralded error  efficiency of  the EG,
and the   electron spins $s_1 s_2$, in this case, could be directly
used in the  recycling  EG  procedure.

In a word,  one can obtain two kinds of useful results with our EG setup.
When only one probe photon is exploited,  the probabilities of heralded success or failure of
the EG are  $\eta_H$ or $\eta_V$, respectively. When the heralded error of EG takes place, a
$|V\rangle$ polarized photon is detected and the state of the spin
subsystem has not been changed. One can input another  probe photon
\emph{p$'$} in state
$|\Phi\rangle_{p'}=\frac{1}{\sqrt{2}}(|R\rangle-|L\rangle)_{p'}$
to repeat the EG process until a horizonal photon $|H\rangle$ is
detected by $D_3$ or $D_4$. This procedure will project the spin
system $s_1s_2$ into an even-parity subspace or  an odd-parity one
eventually. By taking  the recycling procedure into  account, the
total success probability $\eta_{_S}$ of our error-rejecting EG  is
\begin{eqnarray}   % Eq_11
\eta_{_S}=\frac{|r_1-r_0|^2}{4-|r_1+r_0|^2}, \label{etaS}
\end{eqnarray}
which is state independent, resulting in a more efficient quantum
computing \cite{QIP}. Note that each recycling procedure is
conditioned on a click of either vertical detector  $D_1$ or $D_2$,
and it should be stopped when  photon loss takes place.
Subsequently, one has to reinitialize the spins before performing a
new EG operation on the spins.

\section{Cluster state generation with our EG for measurement-based  one-way quantum computing }

Our error-rejecting EG can be  used directly to implement the
one-way quantum computing \cite{ClusterQD3,Oneway1,Oneway2} based on QDs
embedded in optical cavities. In the following, we demonstrate that
our EG  can be used to construct the two-dimensional (2D) QD
cluster state \cite{ClusterQD1,ClusterQD2,ClusterQD3}, which constitutes the
base of one-way quantum computing on solid-state spins.

Suppose there are $j+1$ QD electron spins $\{s_1, s_2, \dots{},
s_{j}\}$ and $ s_{j+1}$, and  $s_{j+1}$ is initialized to be the
state
$\frac{1}{\sqrt{2}}(|\!\uparrow\rangle_{j+1}-|\!\downarrow\rangle_{j+1})$
and the first $j$ spins are initially in the one-dimensional (1D)
cluster state of the form
\begin{eqnarray}   % Eq_12
|\psi_{j}\rangle &=&
(|\!\uparrow\rangle_1+|\!\downarrow\rangle_1{}\hat{Z}_2)(|\!\uparrow\rangle_2
+|\!\downarrow\rangle_2{}\hat{Z}_3)
\cdots\;\;\;\;\;\;\;\;\;\nonumber\\ && \otimes(|\!\uparrow\rangle_{j-1}
+|\!\downarrow\rangle_{j-1}{}\hat{Z}_j)(|\!\uparrow\rangle_{j}+|\!\downarrow\rangle_{j}),
\label{cluster1dj}
\end{eqnarray}
with the phase flip operator
$\hat{Z}_i=|\!\!\uparrow\rangle_i\langle\,\uparrow\!\!|-|\!\!\downarrow\rangle_i\langle\,\downarrow\!\!|$.
To increase the length of the 1D cluster state,  an error-rejecting
EG for spins $s_j$ and $s_{j+1}$ is applied. When the EG fails, the
state of spin $s_j$ is ambiguous and a state measurement on $s_j$
with basis $ \{|\!\uparrow\rangle, |\!\downarrow\rangle \}$ will
collapse the remaining spins into a 1D cluster state of $j-1$
qubits, with or without a $\hat{Z}_{j-1}$ feedback operation.  When
the EG succeeds in the case that is heralded by the click of photon
detector $D_3$, the $j+1$ spins will be projected into
\begin{eqnarray}    % Eq_13
|\psi'_{j+1}\rangle &=&
(|\!\uparrow\rangle_1+|\!\downarrow\rangle_1{}\hat{Z}_2)(|\!\uparrow\rangle_2
+|\!\downarrow\rangle_2{}\hat{Z}_3)\cdots(|\!\uparrow\rangle_{j-1}
\nonumber\;\;\;\;\\ &&
+|\!\downarrow\rangle_{j-1}{}\hat{Z}_j)(|\!\uparrow\rangle_{j}|\!\uparrow\rangle_{j+1}
+|\!\downarrow\rangle_{j}|\!\downarrow\rangle_{j+1}),
\end{eqnarray}
which could be transformed into the 1D cluster state  similar to
that in Eq. (\ref{cluster1dj}) of length $j+1$  by a Hadamard
operation $\hat{H}_{j}$ [$\hat{H}$ completes the following
transformation:
$|\!\uparrow\rangle\rightarrow\frac{1}{\sqrt{2}}(|\!\uparrow\rangle+|\!\downarrow\rangle)$
and
$|\!\downarrow\rangle\rightarrow\frac{1}{\sqrt{2}}(|\!\uparrow\rangle-|\!\downarrow\rangle)$]
performed on spin $j$.  If the success of the EG for  $s_j$ and
$s_{j+1}$ is signaled by the click of  $D_4$, a local operation
$\hat{H}_{j+1}\hat{X}_{j+1}$  (here the spin-flip operator
$\hat{X}=|\!\uparrow\rangle\langle\,\downarrow\!|+|\!\downarrow\rangle\langle\,\uparrow\!|$)
on spin  $s_{j+1}$ could also evolve the ${j+1}$  spins into the
desired 1D cluster state.

This procedure of cluster growth discussed above  could be used to
generate a larger cluster,  since the efficiency of our
error-rejecting EG $\eta_s>0.5$ and can, in principle, approach
unity. To speed up the cluster generation process, some shorter
clusters could be prepared in parallel and then be connected
together to generate the longer one
\cite{ClusterLinear,ClusterLowP,ClusterLowP1}.  In the following, we
introduce an efficient cluster-connecting proposal. Suppose the two
1D clusters \emph{M} and \emph{N} available are, respectively, of
lengths $m$  and $n$,
\begin{eqnarray}    % Eq_14
|\psi_{m}\rangle \!\!&=&\!\!
(|\!\uparrow_{M}\rangle_1+|\!\downarrow_{M}\rangle_1{}\hat{Z}_{M_2})\cdots(|\!\uparrow_{M}\rangle_{m-1}
\nonumber\\
&&+|\downarrow_{M}\rangle_{m-1}{}\hat{Z}_{M_m})(|\!\uparrow_{M}\rangle_{m}
+|\downarrow_{M}\rangle_{m}),\\
|\psi_{n}\rangle\!\! &=& \!\! (|\!\uparrow_{N}\rangle_1+|\!\downarrow_{N}\rangle_1{}\hat{Z}_{N_2})\cdots(|\!\uparrow_{N}\rangle_{n-1}
\nonumber\\
&&+|\downarrow_{N}\rangle_{n-1}{}\hat{Z}_{N_n})(|\!\uparrow_{N}\rangle_{n}
+|\downarrow_{N}\rangle_{n}).
\label{clustermn}
\end{eqnarray}
Before performing EG  on $M_m$ and $N_{1}$, a phase-flip operation
$\hat{Z}_{M_m}$ is applied on spin $M_m$.  The success of the EG
heralded by the click of photon detector $D_3$ will project the
entire spin system into
\begin{eqnarray}     % Eq_15
|\psi^n_{m}\rangle\! \! &=&\! \!
(|\!\uparrow_{M}\rangle_1+|\!\downarrow_{M}\rangle_1{}\hat{Z}_{M_2})\cdots(|\!\uparrow_{M}\rangle_{m-1}+|\downarrow_{M}\rangle_{m-1}{}
\nonumber\\
&&
\otimes\hat{Z}_{M_m})(|\!\uparrow_{M}\rangle_{m}|\!\uparrow_{N}\rangle_1
+|\!\downarrow_{M}\rangle_{m}|\!\downarrow_{N}\rangle_1{}\hat{Z}_{N_2})\nonumber\\
&&\otimes(|\!\uparrow_{N}\rangle_2
+|\!\downarrow_{N}\rangle_2{}\hat{Z}_{N_3})
\cdots(|\!\uparrow_{N}\rangle_{n-1}
\nonumber\\
&&+|\downarrow_{N}\rangle_{n-1}{}\hat{Z}_{N_n})(|\!\uparrow_{N}\rangle_{n} +|\!\downarrow_{N}\rangle_{n}).
\label{clustermn1}
\end{eqnarray}
An additional  Hadamard operation $\hat{H}$ on $M_m$ will evolve the
spin system into a 1D cluster $|\psi_{m+n}\rangle$  of $m+n$ qubits.
As for the case that the success of the EG is signaled by a click of
detector $D_4$,  a local single-qubit operation $\hat{H}\hat{X}$ on
$M_m$ can also evolve the $m+n$ spins   into the 1D cluster
$|\psi_{m+n}\rangle$.

The cluster-connecting procedure based on parity measurement above
is similar to  that  used in  linear optical quantum computing
\cite{linear95}, whereas both the outcome of the EG operation and
the feedback operations after the EG
 are quite different. It generates a 1D cluster of  $m+n$
qubit, rather than $m+n-1$ in linear optical quantum computing
\cite{linear95} or in previous schemes for solid-state spins
\cite{ClusterLinear}  in which the outcomes of the EG can only lead
to an odd parity and the cluster connecting procedure is completed
by a spin measurement  later. One  can also perform a cross-linking
between linear chains to construct a 2D cluster similar to the
previous schemes
\cite{ClusterLinear,linear95,ClusterLowP,ClusterLowP1}, which means
our EG can be used to complete universal one-way quantum computing
efficiently.

\begin{figure}[!tpb]%[tpb]           %Figure 3
\begin{center}
\includegraphics[width=8.25 cm,angle=0]{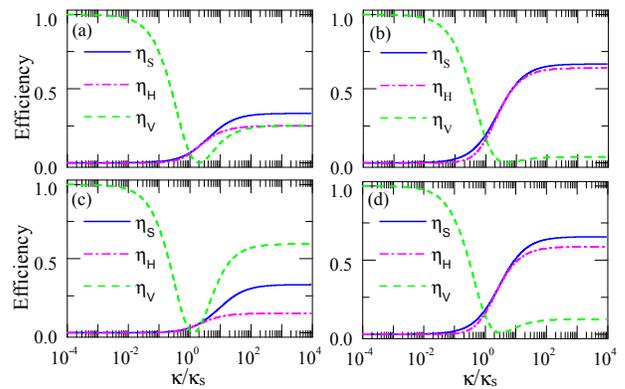}
\caption{The efficiency of the EG  vs  different parameters with
$\omega_{X^-}/\omega_c=1$ and $\gamma/\kappa=0.1$: (a)
$(\omega_c-\omega)/\kappa=0$, $C=1/4$; (b)
$(\omega_c-\omega)/\kappa=0$, $C=1$; (c)
$(\omega_c-\omega)/\kappa=\gamma/\kappa$, $C=1/4$; and (d)
$(\omega_c-\omega)/\kappa=\gamma/\kappa$, $C=1$.} \label{fig3}
\end{center}
\end{figure}

\section{Performance of our error-rejecting EG with current experimental parameters}

The total success probability  $\eta_{_S}$ together with $\eta_{_H}$
and $\eta_{_V}$  of our EG are shown in Fig.~\ref{fig3} as a
function of the side leakage $\kappa/\kappa_s$ with the
cooperativity $C\equiv g^2/\gamma\kappa_{_T}$,
$\kappa_{_T}\equiv\kappa_s+\kappa$, and  $\gamma/\kappa=0.1$
\cite{QDpillar2004}. We tune the transition frequency $\omega_{X^-}$
of the QD to be resonant to that of the cavity,
$\omega_{X^-}/\omega_{c}=1$  \cite{CavityTuning}. When the probe
photon is also resonant to cavity [see Figs.~\ref{fig3} (a) and
(b)], $\eta^r_S=0.255$ and $\eta^p_S=0.559$ can be achieved in the
regime of resonance scattering with $C=1/4$ and the Purcell regime
with $C=1$, respectively, for $\kappa/\kappa_s=13$
\cite{EGQdYoung,CavityLeakage}. When the probe photon detunes from
the  trion transition by $(\omega_c-\omega)/\kappa=\gamma/\kappa$,
shown in Figs.~\ref{fig3} (c) and (d), $\eta^r_S=0.194$ and
$\eta^p_S=0.538$ can be achieved for the same remaining parameters,
and the contribution from the recycling procedure $\eta_{_V}$
increases. Furthermore, the EG could enjoy a higher efficiency with
a lower side leakage and a higher cooperativity $C$, which can be
achieved by utilizing adiabatic cavities with smaller pillar
diameters \cite{QDpillar2004,CavityLeakage}. In other words, the
near -unity efficiency of the error-rejecting EG can be achieved
when the deep Purcell regime with low side leakage is available, and
we can easily attribute this improvement of the efficiency to  the
enhancement of the photon into the cavity mode.

In the above discussion, we can get an efficient error-rejecting EG
for QDs with the perfect spin qubit and the monochromatic
($\delta$-function-like) single photon wavepacket. In fact, every
single photon pulse is of finite linewidth, i.e., a polarized photon
of pulse shape in Gaussian function
$f(\omega)=exp(-\omega/\Delta)^2/(\sqrt{\pi}\Delta)$  with bandwidth
$\Delta$. This finite-linewidth character usually introduces some
additional infidelity in the previous EG protocols
\cite{EGQdHu1,EGQdHu2,QCwei1,QCwei2,QDweak1,QDweak2,QDweak3,EGQdKoshino,EGQdYoung},
while it has little harmful effect on the fidelity of our EG. When
one constitutes our EG with a polarized single-photon pulse $p$ of
Gaussian shape, $|\psi_p\rangle=\frac{1}{\sqrt{2}}\int{}d\omega
f(\omega)[\hat{a}^{\dagger}_R(\omega)-\hat{a}^{\dagger}_L(\omega)]
|0\rangle$, where $\hat{a}^{\dagger}_k(\omega)$ is the creation
operator of a $ k$-polarized photon with frequency $\omega$, the
state of the hybrid system composed of the photon $p$ and electron
spins $s_1$ and $s_2$  just before photon detection process, shown
in Eq. (\ref{pcdfaith}), will be modified to
\begin{eqnarray}    % Eq_17
|\Phi\rangle_{H_2}\!\!&=&\!\!\frac{1}{2}\int{}d{}\omega[r_1(\omega)\!-\!r_0(\omega)]f(\omega)
\Big\{\!(\alpha_1\alpha_2|\!\uparrow\rangle_{s_1}|\!\uparrow\rangle_{s_2}\nonumber\\
&&\!\!-\beta_1\beta_2|\!\downarrow\rangle_{s_1}|\!\downarrow\rangle_{s_2}\!)\otimes{}\hat{a}^{\dagger}_H(\omega)|0\rangle_{p_3}
\!\!+\!\!(\alpha_1\beta_2|\!\uparrow\rangle_{s_1}|\!\downarrow\rangle_{s_2}\nonumber\\
&&\!\!-\beta_1\alpha_2|\!\downarrow\rangle_{s_1}|\!\uparrow\rangle_{s_2}\!)\otimes{}\hat{a}^{\dagger}_H(\omega)|0\rangle_{p_4}\Big\},
\label{pcdfaithm}
\end{eqnarray}
where $
\hat{a}^{\dagger}_H(\omega)=\frac{1}{\sqrt{2}}[\hat{a}^{\dagger}_R(\omega)+\hat{a}^{\dagger}_L(\omega)]$.
Upon the click of photon detector $D_3$ or $D_4$, one can still
complete the EG by projecting spins $s_1$ and $s_2$ into a subspace
of determined parity, as one does with a monochromatic photon wave
packet.  One can find that the fidelity of our EG is independent of
the finite linewidth of the photon pulse, since the
frequency-dependent reflection coefficients $r_j(\omega)$ appear
only in the global coefficient.

In fact,  the effects of dephasing and decay of   electron spins
will affect the performance of the EG. The time needed for the
coherent  control  of single electron spin in QDs is on the scale of
picoseconds \cite{qdop,time} and the cavity photon time is tens of
picoseconds when  the cavity $Q$-factor is about  $ 1\times 10^4
-1\times 10^5$ \cite{QDpillar2004}. Therefore, it is the spontaneous
emission lifetime of a QD, which is about $1$  ns, that sets the
upper limit for the fidelity of the EG.   Meanwhile, the electron
spin coherence time of $10$ ns has been achieved at zero magnetic
field \cite{qd10ns}, and it could be extended to several
microseconds  if the all-optical spin echo technique is exploited
\cite{qd26,QDCspins}. The ratio of the decoherence time of  electron
spins to the operation time needed to complete the EG can exceed
$1\times 10^3$, and thus the fidelity of the EG proposal will be
larger than $0.99$ when taking into account the dephasing process of
the electron spin, which suggests the strong promise of electron
spin in QDs for  scalable quantum computing.

\section{Discussion and summary}

Our scheme of error-rejecting EG can work efficiently with almost
unity fidelity in the strong coupling regime, $g>\kappa_T, \gamma$,
the Purcell regime, $C>1/4$, or even the resonantly scattering
regime, $C=1/4$. It is robust to the imperfections involved in the
practical input-output process, i.e., the nonzero bandwidth,  QD or
cavity decay, and the finite coupling $g/\kappa$, since the fidelity
of our EG is  independent of the reflective coefficients
$r_j(\omega)$ and thus independent of the cooperativity $C$, which
is far different from other schemes that depends on $C$
\cite{EGQdHu2,QCwei1,QCwei2,QDweak1,EGQdYoung,EGQdKoshino}. The
original low fidelity or error items originating from the practical
input-output process are converted into a relatively lower
efficiency in our  EG. Fortunately, the low fidelity or error items
trigger  the single-photon detector $D_1$ or $D_2$, which can be
used to  improve the efficiency of the EG by introducing the
recycling procedure. In fact, our recycling procedure can contribute
little when perfect circular birefringence ($C\gg1$ and
$\omega_{X^{\!\!-}}$$=\omega_c=\omega$ ) is available, since the
efficiency $\eta_H$ of the  EG  without recycling procedure
approaches unity in this situation. Although our proposal is
detailed with the QD-cavity system, it could also be implemented
with solid-state spin coupled to a photonic crystal waveguide
\cite{spinwaveguide}.

The previous EG performed in a resonantly scattering regime, $C=1/4$
and $\omega_{X^{\!\!-}}$$=\omega_c=\omega$ could also be completed
with high fidelity, since a reflectivity $r_1=0$ in the resonantly
scattering regime could be automatically eliminated, and only the
photon that decouples the electron spin  could be reflected.
Therefore, one can entangle  two spins by subsequently probing the
two spin-cavity systems with a linear polarized photon or entangle
two linear polarized photons by subsequently importing them into a
spin-cavity system, where the even-parity subspace of  the spin
system or the photon system could be easily picked out, since the
odd-parity case will inevitably be signaled by photon loss
\cite{EGQdYoung}, and the corresponding efficiency of the EG is
$0.25$. It is quite different from our EG where the linear polarized
photon, after being reflected by the QD-cavity system, in the ideal
case $C\gg1$ and $\omega_{X^{\!\!-}}$$=\omega_c=\omega$,  is
supposed to change its polarization into the orthogonal polarization
and exert a phase-flip operation on the spin. The interference of
the photon after being reflected by two cavities in superposition
can project the two spins into either even-parity subspace or
odd-parity subspace in a heralded way.

The  error-rejecting EG only involves one effective input-output
process, which makes our scheme  more efficient than others since
the practical input-output coupling  $\eta_{in}<1$
\cite{cavitypillarcoupling}. In this situation, the probe  photon
can be reflected directly by the cavity, and it is harmful and will
reduce the fidelity of the entangling process in the other schemes
\cite{EGInter5,EGInter6,EGQdHu2,QCwei1,QCwei2,QDweak1,EGQdYoung,EGQdKoshino}.
However, it can only lead to a  decrease of the efficiency of our
EG, since the state of the photon reflected directly by the
microcavity together with that of the spins will be  kept unchanged.
In other words, the photon reflected directly by the microcavity  is
still in $|V\rangle$ polarization and it  will trigger the
single-photon detector $D_1$ or $D_2$, which signals the restarting
of the  EG. This  makes the EG different from the one based on a
double-sided  cavity where the photon is encoded in its Fock state
\cite{NVNemoto}. The photon loss during the EG process owing to the
inefficiency of the single-photon detector or cavity absorbtion will
decrease the efficiency of the EG, but it does not affect the
fidelity of our EG since both the success of the EG and the
restarting of the EG are signaled by a click of  single photon
detectors.

In conclusion, we have proposed  an efficient error-rejecting  EG
proposal  for two electron spins of QDs  embedded in low-$Q$ optical
microcavities.  With our error-rejecting EGs,  a cluster-state
connection scheme could be completed efficiently. Under the
practical experimental condition, the EG could be performed well
with  almost unity fidelity and an efficiency of $\eta_s>0.53$ for
$C=1$. We believe the EG could provide a promising building block
for solid-state scalable quantum computing and quantum networks in
the future.

\section*{ACKNOWLEDGMENTS}

This work is supported by the National Natural Science Foundation of
China under Grants No. 11474026 and No. 11674033, and the
Fundamental Research Funds for the Central Universities under Grant
No. 2015KJJCA01.

\end{document}